# The Big Bang Problems: Anisotropy of z ≤ 6 Redshifts

A.V. Glushkov[*]

Yu.G. Shafer Institute of Cosmophysical Research and Aeronomy,

31 Lenin Ave., 677980 Yakutsk, Russia

## Abstract

The three-dimensional space distribution of 48921 quasars and 16113 Seyfert galaxies with redshifts $z \leq 6$ is investigated. The global anisotropy caused by the shift of the observer place by $\approx$ 50 Mpc from a center of their symmetry (supposed center of the Metagalaxy) to the side of the vector with equatorial coordinates $\alpha \approx 13°$ and $\delta \approx 70°$ has been found in the placement of these objects. In the opposite direction there exists the extensive region where the progressive decrease of redshifts up to a minimum (near $\alpha \approx 193°$, $\delta \approx -70°$) is observed. The influence of gravitational potential and possible rotation of the Metagalaxy on the anisotropy of redshifts of the cosmological objects has been considered.

## 1. Introduction

The majority of modern cosmological models are based on the assumption that all points of space are equivalent; therefore, the Universe is on average uniform and isotropic. This hypothesis is confirmed by the observations of isotropic cosmic microwave background radiation [1]. However, the distribution of matter is extremely nonuniform in the close vicinity of the Galaxy; moreover, there is a distinct hierarchy on a larger scale (from a few hundred kpc to a few hundred Mpc). Galaxies enter into the composition of clusters and superclusters, which in turn form a cellular structure of the Universe, with a characteristic size of nonhomogeneities being 100 to 130 Mpc (see, for example, [2, 3]). The cellular structure of the Universe was reliably established on the basis of statistical analysis of the distribution of galaxies for moderate redshift values of $z \leq$ 0.5. Cosmic rays of ultrahigh energy also confirm the existence of such a structure [4–8].

The ladder of the macrocosmos does not end with a cellular structure of the substance of sizes mentioned above. A periodicity in the argument $\log(1 + z)$ was discovered in the distribution of quasars [9]; later on, this periodicity was repeatedly confirmed and refined (see, for example, [10, 11]). In the spectra of quasars, Ryabinkov *et al.* [9] observed cosmological variations of the spacetime distribution of 847 absorptive systems. These variations manifest themselves as a successive appearance of maxima and minima of functions in the arguments $\ln(1 + z)$ and $(1 + z)^{-1/2}$. It is believed that this structure is due to the alternation of distinguished and depressive eras in the course of a cosmological evolution that are separated by a characteristic time interval in the range $520 \pm 160$ million years, its specific duration being dependent on the choice of cosmological mo-del. In [8] the global anisotropy of spatial distribution of quasars, which is characterized by their central asymmetry, was found.

---

[*] e-mail: a.v.glushkov@ikfia.ysn.ru

Quasars are the most puzzling objects in the Universe and the most powerful sources of radiation there. For the overwhelming majority of quasars, the redshifts exceed 0.1, but, for some of them, the redshifts are as large as about 5 to 6 [12]! This suggests that they occur at cosmological distances exceeding many hundred megaparsecs. It is assumed that quasars characterised by the largest redshifts are the most remote from us and belong to the group of the very first galaxies formed upon the beginning of the Big Bang in the expanding Universe [1].

Below are given some new results of research of spatial distribution of quasars and Seyfert galaxies which throw the additional light on a problem of origin of the Universe.

## 2. Anisotropy of Quasars

Before to begin the analysis of the data, we shall consider three spherical coordinate systems, which we shall use below (Fig. 1). One of them is a galactic system shown by the plane $G$ with the normal vector of North Pole $N_G$, which has equatorial coordinates $\alpha_G = 192.25°$ и $\delta_G = 27.4°$. The Supergalaxy plane (Local supercluster of galaxies) is located almost perpendicular to the Galaxy plane. Its North Pole ($N_{SG}$) has equatorial coordinates $\alpha_{SG} = 286.2°$ и $\delta_{SG} = 14.1°$.

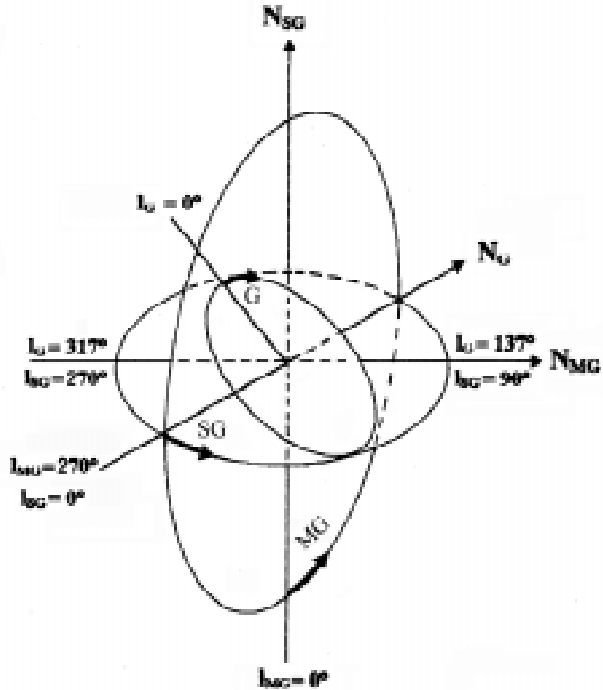

**Fig. 1.** The scheme of mutual arrangement of galactic (*G*), supergalactic (*SG*) and metagalactic (*MG*) of spherical coordinate systems. A normal vector NG, NSG and NMG represents directions of the appropriate North Poles; the arrows represents directions of readout in all three coordinate systems from the appropriate zero marks.

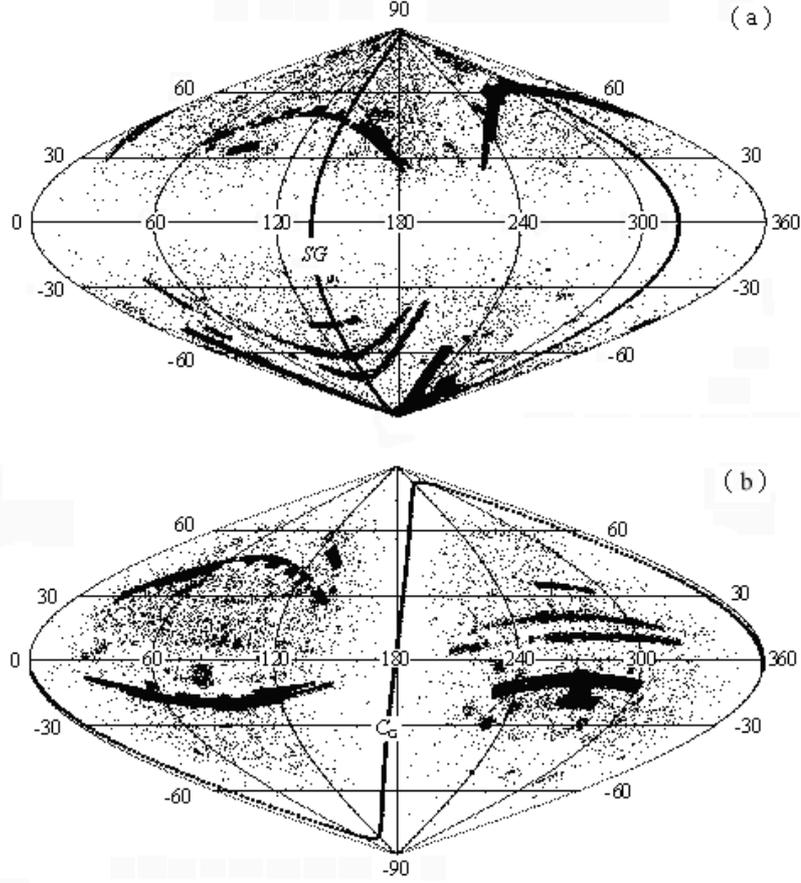

**Fig. 2.** A distribution of 48921 quasars from the catalogue [12] on the developed celestial sphere in galactic (a) and metagalactic (b) coordinates. Curves *SG* and $C_G$ represents the planes of the Supergalaxy and the Galaxy; $C_G$ is the centre of the Galaxy.

Let us now take the line of crossing of planes of the Galaxy and the Supergalaxy with the longitude in galactic coordinates $l_G = 137°$ (in supergalactic coordinates $l_{SG} = 90°$) as the North Pole of one more plane *MG*, convenient for our analysis, which we shall name conditionally as a metagalactic one. Its normal (vector $N_{MG}$) has the equatorial coordinates $\alpha_{MG} = 40.6°$ и $\delta_{MG} = 59.5°$. The longitude in all three systems of coordinates is counted in clockwise direction from the appropriate zero marks.

In [8] the sample of 3339 quasars with redshifts $z \leq 2.5$ of the catalogue [13] has been considered. This sample has allowed us to draw a sufficiently correct picture of spatial distribution of quasars and to specify a possible existence of the Metagalaxy. However, such a conclusion based on the relatively small number of objects, needs the additional confirmation. Therefore, we use here the most complete version of the periodically updated catalogue of galaxies with active nuclei [12] containing 48921 quasars and 16113 Seyfert galaxies.

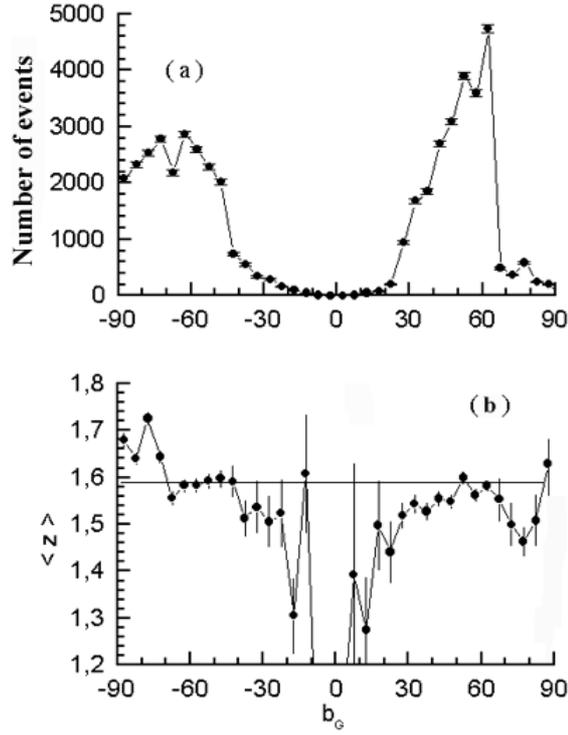

**Fig. 3.** Distributions of a number of quasars (a), submitted in Fig. 2a, and their average values redshifts (b) depending on the galactic latitude (with a step of $\Delta b_G = 5°$). The horizontal line represents the average value.

At first let's consider some general characteristics of the spatial distribution of quasars. In Fig. 2a the map of the quasars with $z \leq 6$ from the mentioned catalogue on the developed celestial sphere in galactic coordinates is shown. The curve *SG* represents the disk of the Supergalaxy. In separate places the local areas with a relatively high density of quasars are visible. They are caused, mainly, by a technique of making catalogue. The catalogue is replenished at the expense of deep observation of the sky on the small platforms chosen separately.

### 2.1. Northern-Southern Asymmetry in Galactic Coordinates

Figure 3a shows the distributions of a number of quasars submitted in Fig. 2a, depending on a galactic latitude (with a step of $\Delta b_G = 5°$). In northern and southern hemispheres there are 23928 and 24993 objects, respectively. Near the equatorial area of the Galaxy because of a strong absorption of the light there are only few of them. In Fig. 3b the distribution of average redshifts in such a representation of the data is shown. The horizontal line corresponds to $<z>_Q = 1.586 \pm 0.003$ for the whole catalogue. Here the latitudinal dependence manifests itself much more poorly. It is noticeable only in the narrow area of the Galaxy disk i.e. in the band of latitudes $|b_G| \leq 20°$.

In Fig. 3b the northern-southern asymmetry of redshifts pays attention to itself, in northern and southern hemispheres $<z>_N = 1.558 \pm 0.004$ and $<z>_S = 1.614 \pm 0.004$. The difference $\Delta z = <z>_S - <z>_N \approx 0.056$ exceeds the casual deviation (for two roughly similar samples in a number of events) by $0.056/0.004 \approx 14\sigma$, that is the extremely improbable event

If we use the Hubble law

$$r = v/H = \beta R_M \text{ [Mpc]}, \qquad (1)$$

where $r$ is the distance to the object being considered, $\beta = v/c$ (here, $v$ and $c$ are the velocity of the object and the velocity of light, respectively), and $H \approx 75$ km/s Mpc is the Hubble constant, then these results can be treated as the distances $<r>_N$ and $<r>_S$ from the observation point to a "average quasar" in the north-south directions of the Galaxy. The value $R_M \approx cT \approx 10^{26}$ m $\approx 4000$ Mpc characterises the ultimate limiting dimensions of the Metagalaxy ($T \approx 1/H \approx 4 \times 10^{17}$ s $\approx 13$ billion years is the age of the Metagalaxy since the beginning of the Big Bang [1]). In the relativistic case, the relative velocity $\beta$ is related to the Doppler redshift $z$ by the equation [1]

$$\beta = ((1+z)^2 - 1)/((1+z)^2 + 1), \qquad (2)$$

which will be used below in the estimating of distances $r$. From (1) and (2) we find values $<r>_N \approx 2940$ Mpc and $<r>_S \approx 2980$ Mpc, which specify the shift $\Delta R_G \approx 20$ Mpc a point of observation concerning the equator of the sphere with radius $<R> \approx 2960$ Mpc in a northern direction of the Galaxy (see Fig. 1).

### 2.2. Equatorial Asymmetry in Galactic Coordinates

Now let's consider the distribution of quasars in galactic spherical sectors (with a step in longitude of $\Delta l_G = 10°$). The distribution is shown in Fig. 4 for a number of events (a) and average values of redshifts (b). The horizontal line corresponds to average value of $<z>_Q = 1.586$ for the whole sample. The dashed curve represents the approximation of data by the function

$$f(l_G) = f_0 \cdot (1 + A_1 \cdot \cos(l_G - l_1)). \qquad (3)$$

The best values of amplitude $A_1$ and the phase $l_1$ are found by minimizing the value

$$X^2 = \sum_{i=1}^{n} (f_i - N_i)^2/f_i, \qquad (4)$$

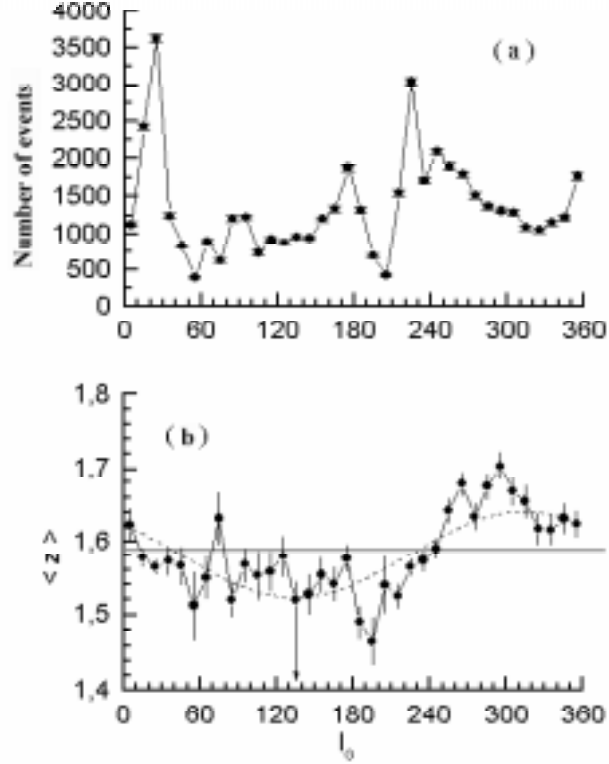

**Fig. 4.** Distributions of a number of quasars (a), submitted in Fig. 2a, and their average redshifts (b) in dependence of the galactic longitude (with a step of $\Delta l_G$ = 10°). The horizontal line corresponds to the average value; the dashed curve represents the best approximation of the data by the function in (3).

where $N_i$ is the number of quasars in the $i$th sector $(\Delta l_G)_i$,

$$f_0 = (\sum_{i=1}^{n} N_i)/n = N/n$$

and

$$A_1 = (f_{max} - f_{min})/(f_{max} + f_{min}) \;. \tag{5}$$

In Fig. 4b the anisotropy of quasars has the more brightly expressed character. Their distribution obviously contradicts to the horizontal line for the isotropic case, with a probability of random outcome $< 10^{-7}$. It is seen, that the data agree much better with approximation (3), which has the minimum at $l_G \approx 136°$.

It is curious to note, that extremums of function (3) lay in the Supergalaxy plane which is crossed almost perpendicularly with the Galaxy plane along the line with $l_G \approx 137.4°$ and $317.4°$ (see Fig. 1 and Fig. 2a). The average values of redshifts in the specified directions are equal $<z>_{min} \approx 1.52$ and $<z>_{max} \approx 1.64$. The distances up to "average quasars" $<r>_{min} \approx 2910$ Mpc and $<r>_{max} \approx 2998$ Mpc correspond to them, which indicate to the shift $\Delta R_{MG} \approx 44$ Mpc of the point of observation relative to the equator of sphere with the radius $<R> \approx 2954$ Mpc, along the axis $N_{MG}$ (see Fig. 1).

### 2.3. Northern-Southern Asymmetry in Supergalactic Coordinates

Above we have found out the shifts of point of our position relative to the centre of sphere of "average quasars" along $N_G$ and $N_{MG}$ axes on $\Delta R_G \approx 20$ Mpc and $\Delta R_{MG} \approx 44$ Mpc. Let's consider additionally $<z>$ in the northern-southern direction of the Supergalaxy, which will allow to reveal a possible asymmetry of quasars in the direction of the third axis $N_{SG}$ (see Fig. 1). In Fig. 5a the distribution of the number of quasars is shown depending on the supergalactic latitude (with a step of $\Delta b_{SG} = 5°$). In northern and southern hemispheres there are 21935 and 26986 objects, respecti-vely. Near the poles of the Supergalaxy because of a strong absorption of light in the equatorial region of the Galaxy they are almost absent. In Fig. 5b the distribution of average redshifts is shown. The horizontal line corresponds to $<z>_Q$. Here the latitudinal dependence relative to the plane of the Supergalaxy manifests itself more poorly, than in Fig. 3b. In northern and southern hemispheres we have $<z>_N = 1.572 \pm 0.005$ and $<z>_S = 1.596 \pm 0.004$. From (1) and (2) we find values $<r>_N \approx 2949$ Mpc and $<r>_S \approx 2967$ Mpc, which indicates to the shift $\Delta R_{SG} \approx 9$ Mpc of a point of observation relative to the equator of sphere with the radius $<R> \approx 2958$ Mpc in a northern direction of the Supergalaxy (Fig. 1).

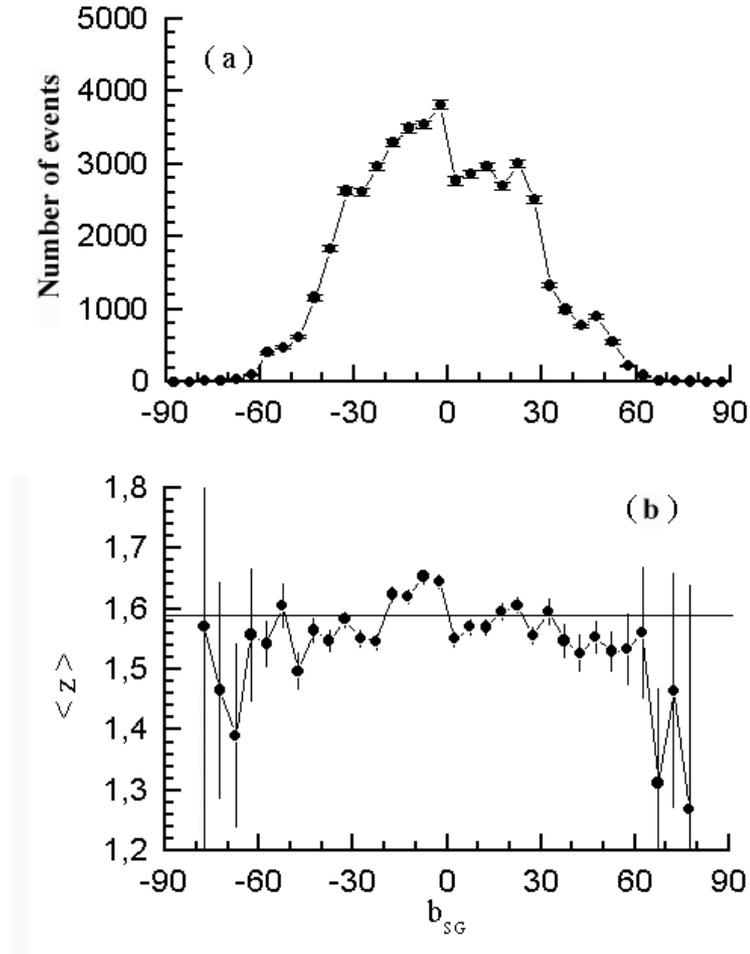

**Fig. 5.** Distributions of the number of quasars (a) and their average redshifts (b) from the catalogue [12] depending on a supergalactic latitude (with a step of $\Delta b_{SG} = 5°$). The horizontal line corresponds to average value.

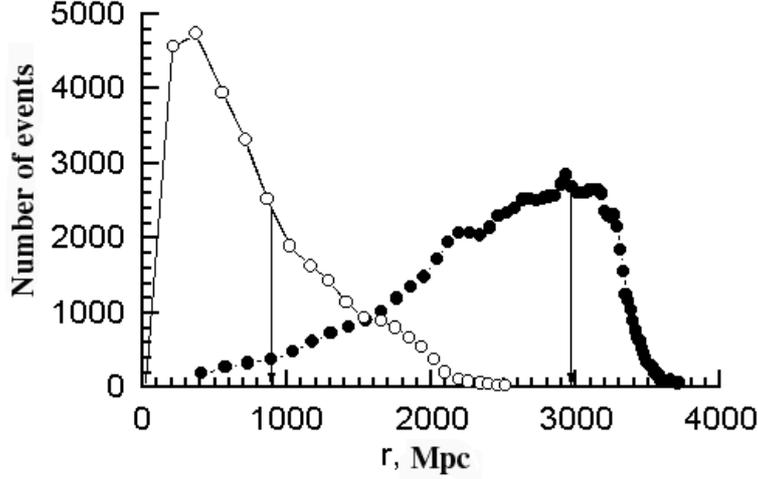

**Fig. 6.** Distribution of the number of Seyfert galaxies (o) and quasars (•) from the catalogue [12] depending on their distance (1) up to the observer. The arrows represent the radii of spheres of "average quasars" and "average Seyferts".

### 2.4. Central Asymmetry

Figure 6 shows the distribution of the number of quasars at different distances from the supposed centre of symmetry of their volumetric configuration. As the shifts $\Delta R_G$, $\Delta R_{SG}$ and $\Delta R_{MG}$ are mutually perpendicular (see Fig. 1), then in this case the shift of observation point relative to this centre is equal to

$$\Delta R = \sqrt{(\Delta R_G)^2 + (\Delta R_{SG})^2 + (\Delta R_{MG})^2} \approx 50 \text{ Mpc}. \qquad (6)$$

This point has the Supergalaxy coordinates $l_{SG} \approx 113° \pm 9°$ and $b_{SG} \approx 9° \pm 4°$ (the correspondent equatorial coordinates are $\alpha \approx 65°$ и $\delta \approx 83°$). The sphere of "average quasars" with the radius $<R> \approx 2960$ Mpc is marked by the arrow in Fig. 6. The results in Fig. 3b–5b say that we are not in the centre of this sphere, and are shifted from it by $\approx 50$ Mpc in the direction of the vector with the coordinates $l_{SG} \approx 113°$ and $b_{SG} \approx 9°$. In the framework of the hypothesis [8] on the Metagalaxy, a direction opposite to the mentioned one ($l_{SG} \approx 293°$, $b_{SG} \approx -9°$) indicates to the centre of the Metagalaxy.

## *2.5. Topology of Redshifts*

Let us now analyze, in addition, a spatial distribution of quasars by another method. Let's consider their average redshifts in the circle with the radius $35°$, whose centre consistently moves in right ascension by $1°$ (from $0°$ up to $360°$) and bypasses all declinations (from $-90°$ up to $90°$) through $1°$. Such a circle completely blocks all emptiness in Fig. 2 and enables to receive the detailed picture of distribution $<z>$ on the whole sphere.

In Fig. 7a, the deviations

$$n_z = (<z>_{\alpha\delta} - <z>_Q)/\Delta z \qquad (7)$$

of the observable average redshifts $<z>_{\alpha\delta}$ in the above mentioned circle with coordinates of its centre $\alpha$ and $\delta$ from $<z>_Q = 1.586$ for the whole sample are shown. The received size $n_z$ was attributed to a platform of the $1°\times1°$ size, and platform itself consistently moved on the whole sky. The limits of change of $n_z$ are specified on the bottom of Figure as the shaded scale with a step of $\Delta z = 0.08$. The extreme values of $n_z = \pm 3$ refer to $<z> = 1.826$ and $1.346$, which correspond, proceeding from equations (1) and (2), to distances $\approx 3120$ Mpc and $\approx 2780$ Mpc up to the farthest and nearest points of the surface of "average quasars". In order the presentation to be clearer, the equatorial coordinates are also given here. The curve *SG* represents the disk of the Supergalaxy.

The map of Fig. 7a confirms the presence of global anisotropy in the distribution of quasars by their remoteness from the observer. The anisotropy directions found above (chapter 2.4) are shown by dark and light circles (with coordinates $\alpha \approx 65°$, $\delta \approx 83°$ and $\alpha \approx 245°$, $\delta \approx -83°$) near the northern and southern poles of the Earth. However, in Fig. 7 there are some new structural features of redshifts. Among them there are two "black holes" located in the sphere points opposite from each other, near the places of visible crossing of planes of the Galaxy and the Supergalaxy which attract a special attention to themselves. This local topological inhomogeneties of redshifts are not connected with a small number of quasars in the mentioned areas of the sky, because similar "holes" are not present in other places near the equator of the Galaxy (for example, near the centre of the Galaxy, where the light is absorbed most strongly). We believe, that the occurrence of these "holes" is not accident, and caused, probably, by some physical reasons.

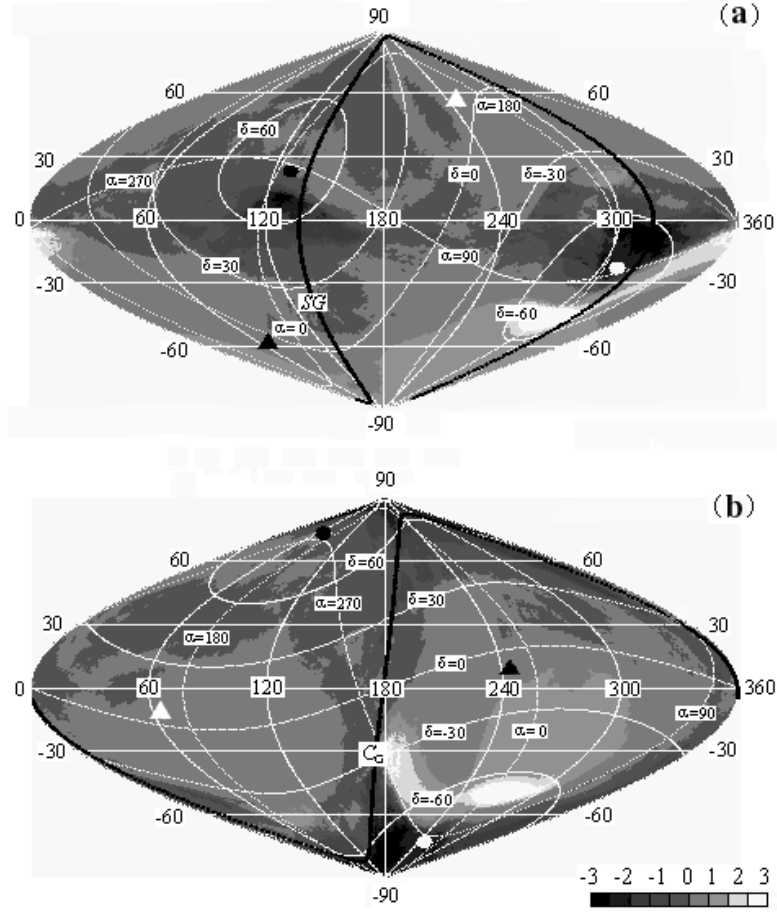

**Fig. 7.** Deviations of local redshifts from $<z>_Q = 1.586$ of the whole samples of quasars from the catalogue [12] in galactic (a) and metagalactic (b) coordinates (submitted in Fig. 2). The shaded scale represents limits of changes of value (7) with a step of $\Delta z = 0.08$; the curves $SG$ and $C_G$ correspond to the planes of the Supergalaxy and the Galaxy; $C_G$ represents the centre of the Galaxy; the light and dark circles correspond to the directions of shift of the observer from the supposed centre of sphere of "average quasars" and the direction to the nearest point of this sphere respectively; the light and dark triangles correspond to the directions of the maximum deviation of temperature of cosmic microwave background radiation from an average level all sky ($\approx 2.7$ K) by $+3.5$ and $-3.5$ mK respectively [1].

Let's now try to understand it more deeply. For convenience of the further analysis we shall turn to another coordinate system, similar to the metagalactic one. Let's choose its northern and southern poles in the direction of vectors with coordinates $\alpha_N = 13°$, $\delta_N = 70°$ and $\alpha_S = 193°$, $\delta_S = -70°$ respectively, approximately coinciding with the centres of both "black holes". In this case the distribution of quasars from the catalogue [12] on the developed celestial sphere looks like as is shown in Fig. 2b. And the distribution of values (7), describing the deviation of local average redshifts from $<z>_Q = 1.586$ of the whole sample, is shown in Fig. 7b.

It is remarkable that the extremely non-uniform distribution of quasars in Fig. 2b has nevertheless a rather symmetric distribution of average redshifts in Fig. 7b, where the deviations of distances from the radius $<r>_Q \approx 2960$ Mpc of the sphere of "average quasars" into both sides do not exceed $\approx 170$ Mpc. Along the disk of the Galaxy in Fig. 7b the topological hollow is clearly seen, probably caused by a technique of drawing up of the catalogue [13], where the farthest quasars are not simply visible because of the strongest absorption of light.

Nevertheless, the "black holes" marked above remain near the poles in Fig. 7b. They have, in my opinion, the independent nature which is not connected with the number of objects in the catalogue [12]. We already observed them clearly in the catalogue [13], where there were 3594 objects. The zones of manifestation of these "holes" in environmental areas of the sky are much larger than the width of hollow near the equator of the Galaxy. One can judge about it by the change of average redshifts of quasars near the poles in Fig. 8 (closed circles), where the fast decrease $<z>$ up to 0 as they approach the poles in Fig. 2b is clearly visible. And the "hole" near the South Pole is expressed, apparently, more strongly in comparison with the northern direction.

The open circles in Fig. 8 show values

$$<n_z> = (\sum_{i=1}^{k}(n_z)_i)/k ,\qquad(8)$$

received by averaging of all $n_z$ in Fig. 7b in intervals of $\Delta b_{MG} = 5°$. These values change more smoothly because of strong smoothing $<z>$ in Fig. 7.

## *2.6. Modeling*

The better to understand the physical sense of the results received above, we shall check up our hypothesis about a global central asymmetry of three-dimensional spatial distribution of "average quasars" on the mathematical model. Let's calculate at first the distances up to the surface of sphere with the radius 2960 Mpc from the point shifted from the centre of sphere by 50 Mpc in the direction of vector with the equatorial coordinates $\alpha_N \approx 13°$ и $\delta_N \approx 70°$. Then we shall find redshifts, interesting for us, from these distances by back recalculation using the formulae (1) and (2). Let's take into account also the influence of absorption of light in the equatorial region of the Galaxy on average redshifts in this region. For this purpose let's multiply all calculated values of z by the coefficient

$$\eta_1 = 1 - \exp(-b_G/9) ,\qquad(9)$$

chosen by us empirically. Besides that let's eliminate from the analysis all events in the band latitudes $|b_G| \leq 15°$, where there are almost no data of observation

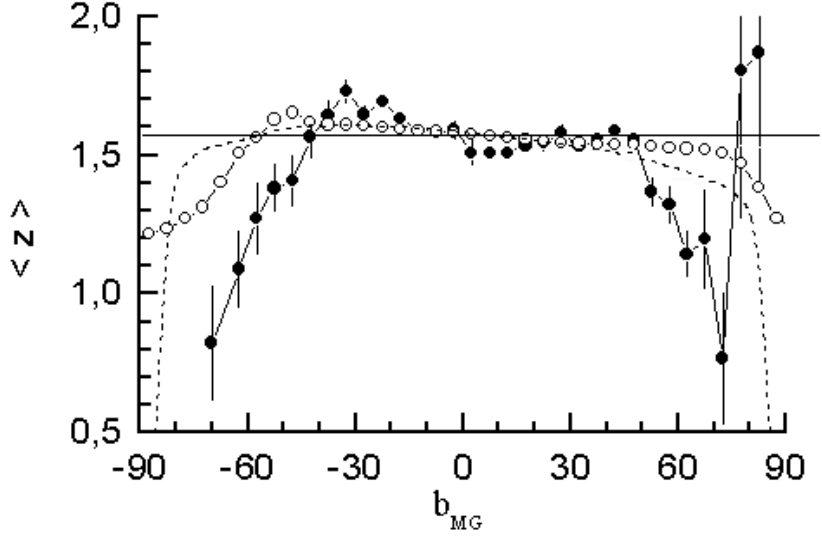

**Fig. 8.** Distributions of average redshifts of quasars, submitted in Fig. 2b, (closed circles) and average sizes (7), submitted on Fig. 7b and Fig. 9b, (open circles and dashed curve respectively) depending from the latitude of the Metagalaxy (with a step of $\Delta b_{MG} = 5°$). The horizontal line represents the average value.

The average redshifts (in the coordinate system of Fig. 2b) calculated in such a way are shown in Fig. 8 by a dashed curve, and the topological map of their values (7) are in Fig. 9a. It is seen that the dashed curve in the field of latitudes $|b_{MG}| \leq 40°$ does not roughly contradict to closed circles and agrees well with the open circles reflecting the latitudinal dependence of smoothed redshifts in Fig. 7b. However, in polar areas (at $|b_{MG}| > 40°$) there is no such a good agreement. Here some additional change of the model is required to result it in more complete agreement with results of observation.

From comparison of Fig. 7b and Fig. 9a it is seen, that they are very similar between themselves on the whole, except regions near the poles, where in Fig. 7b the "black holes" mentioned above are clearly marked out. The calculations have shown, that such defects on the surface of sphere of "average quasars" are possible only at the expense of the fast decrease of redshifts caused by something around "holes" as they approach the centre. We have considered one of the possible variants of formal decision of this task by the additional introduction of linear factor

$$\eta_2 = |b_{MG} - b_0|/20 \qquad (10)$$

with parameters $b_0 = 13°$ and $17°$ for northern and southern poles respectively, by which we have additionally multiplied the rated values $z_1 = \eta_1 z$ at $b_0 < |b_{MG}| \leq 35°$, and other events around poles (in circles with radii $b_0$) have been excluded from the analysis. The results of modeling are shown in Fig. 9b. Here there is a sufficiently close similarity with the map of Fig. 7b. Possible mechanisms of formation of the "black holes" from the physical point of view will be discussed below.

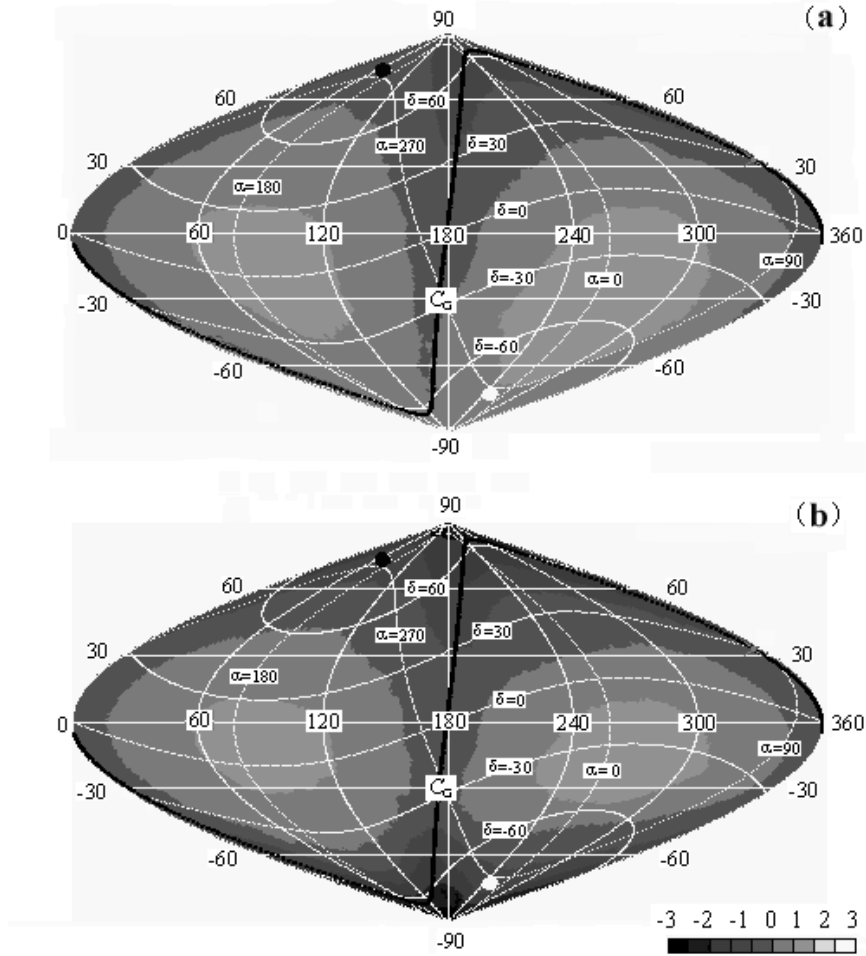

**Fig. 9.** Expected values (in metagalactic coordinates) of average redshifts of quasars calculated by the formulae (1) and (2) for distances up to the surface of sphere with the radius 2960 Mpc from the point shifted from the centre of sphere by 50 Mpc toward the vector with equatorial coordinates $\alpha_N = 13°$ and $\delta_N = 70°$. Dark curves represents the plane of the Galaxy; $C_G$ is its centre; (a) shows the decrease of $z$ in the disk of the Galaxy because of the absorption of light; (b) shows local attenuation (10) of redshifts near to poles of the Metagalaxy in addition to (a); light and dark circles – see Fig. 7.

## 3. Anisotropy of Seyfert Galaxies

Let's now consider one more class of objects which are similar with quasars in many respects i. e. Seyfert galaxies (Seyferts). They have considerably smaller values of redshifts and, therefore they, are much closer to us. It is supposed that these objects have appeared at a later sta-ge of evolution of a substance in the extending Universe [1]. Our interest to Seyfert galaxies is caused by a desire to be convinced once again of the real existence of global anisotropy of the substance in a cosmological scale on the basis of other data.

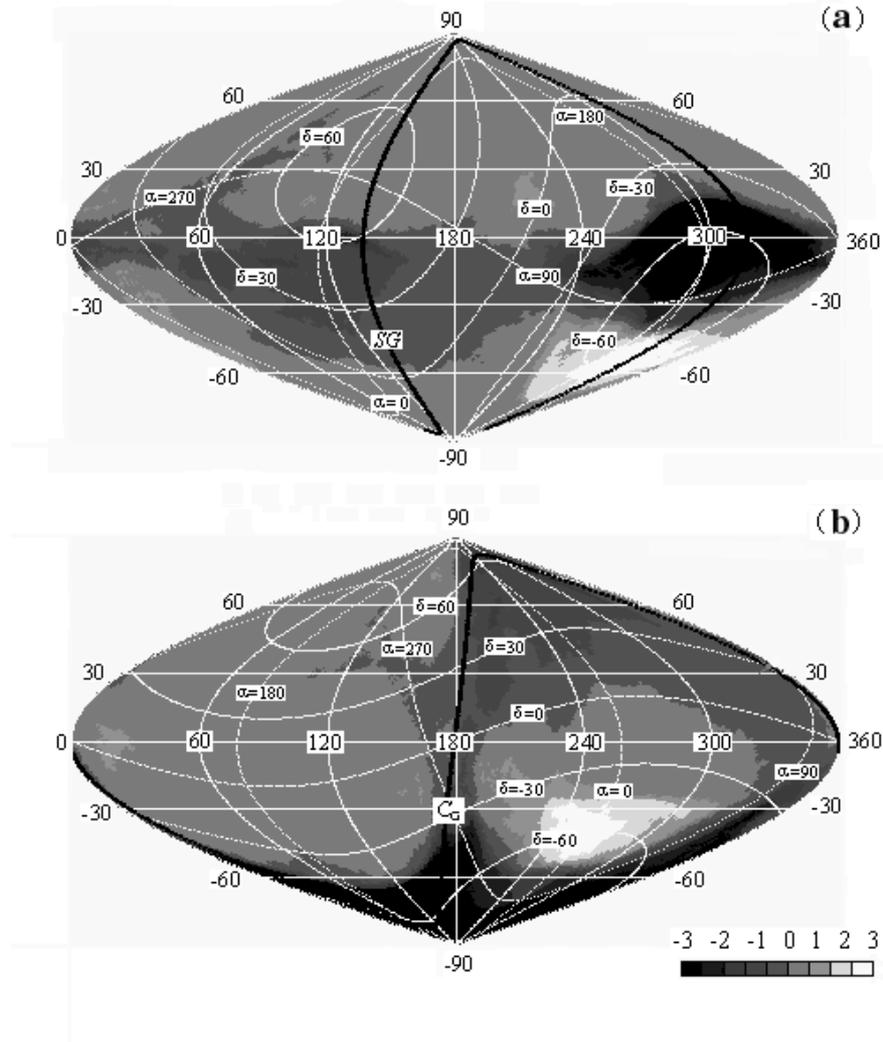

**Fig. 10.** Deviations of local redshifts of Seyfert galaxies from $<z>_{Sy} = 0.16$ of the whole sample from the catalogue [12] in galactic (a) and metagalactic (b) coordinates. The shaded scale represents the limits of changes of the value (7) with a step of $\Delta z = 0.015$; dark curves $SG$ and $C_G$ correspond to the planes of the Supergalaxy and the Galaxy; $C_G$ is the centre of the Galaxy.

In Fig. 6 the open circles show the distribution of 16113 Seyfert galaxies [12] depending on their remoteness from the observer. They have been found using the formulae (1) and (2). An average redshifts of the whole sample is equal to $<z> = 0.253 \pm 0.002$. It corresponds to the sphere of "average Seyferts" with the radius $\approx 890$ Mpc marked in Fig. 6 by the arrow. It is seen, that Seyfert galaxies and quasars form two spatial structures, with the maxima of distributions at $\approx 350$ Mpc and $\approx 2960$ Mpc. In the range of $\approx 1400-1800$ Mpc their number is approximately identical, though this range can, probably, slightly vary. We shall consider further not all Seyfert galaxies, only those which are nearest us, at distances ≤ 1400 Mpc. Farther Seyfert galaxies are presented extremely non-uniformly in the catalogue [12]. They can bring in the methodical distortions into a picture interesting for us.

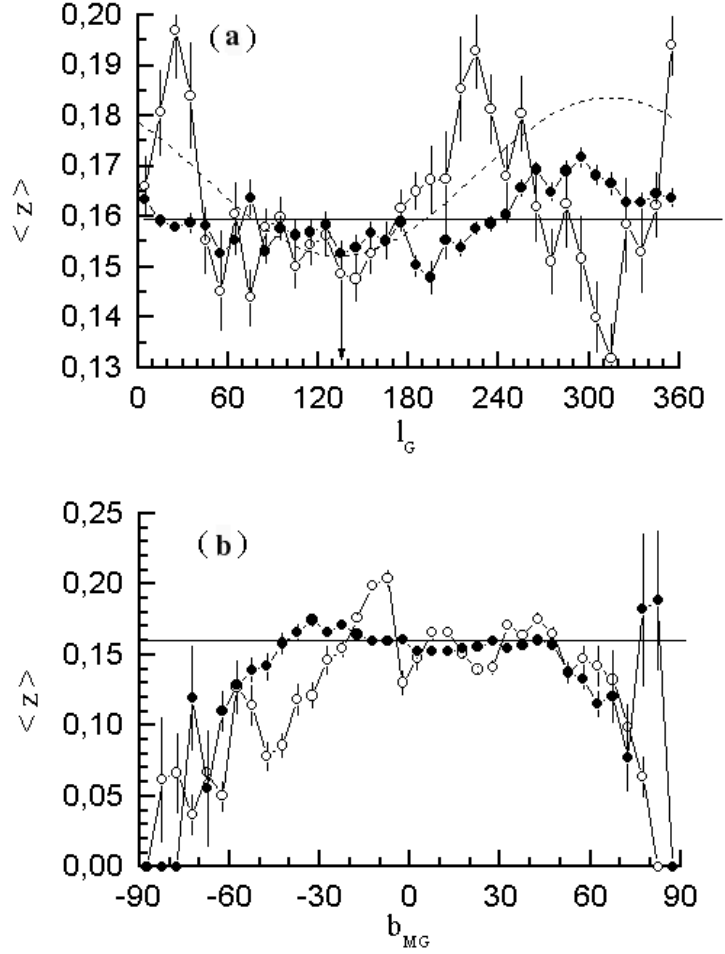

**Fig. 11.** Distributions of average redshifts of seyfert galaxies (○) and of quasars (●) in dependence of a galactic longitude (a) and metagalactic latitude (b). The horizontal lines correspond to average values of $<z>_{Sy} = 0.16$; the dashed curve represents the approximation (3).

Figure 10 shows the map of average redshifts of 12655 objects with $z \leq 0.4$ in galactic (a) and metagalactic (b) coordinates. They show deviations of individual $<z>$ in the scanning circle with the radius 45° from a general value for this sample $<z>_{Sy} = 0.160 \pm 0.001$ in units $\Delta z = 0.015$. The centre of circle (as well as in Fig. 7) has consistently moved with the step 1°. Its sizes have allowed to exclude reliably the appearance of any casual sizes (7). And it, in turn, has given us an opportunity to receive a reliably correct picture of distribution of the average redshifts of Seyfert galaxies on the whole sky.

From the comparison of Fig. 7 and Fig.10 it is seen, that they are similar among themselves in many respects, but the anisotropy of Seyfert galaxies looks even more expressively. Also the distribution itself of average redshifts in Fig. 11a (open circles) says about it where the closed circles show a similar distribution for quasars for comparison in Fig. 4b. The last one has been decreased as abscissa axis by $<z>_{Sy}/<z>_Q = 0.16/1.586 \approx 10$ times. The horizontal line corresponds to $<z>_{Sy} = 0.16$.

A huge "black hole" around South Pole in Fig. 10b has led to an appreciable failure (with the minimum at $l_G \approx 315°$) in distribution of average redshifts of Seyfert galaxies in Fig. 11a. If we exclude it from the analysis ((remove all $<z>$ in the sector $\Delta l_G \approx 270° - 350°$), then in this case data do not contradict to the approximation (3) with the minimum at $l_G \approx 135°$ (dashed curve). In the minimum and the maximum of this function we have $<z>_{min} \approx 0.152$ and $<z>_{max} \approx 0.183$. Distances up to "average Seyferts" $<r>_{min} \approx 572$ Mpc and $<r>_{max} \approx 673$ Mpc correspond to them, which are still indicative of the shift $\Delta R_{MG} \approx 50$ Mpc of the point of observer relative to the of "average Seyferts" sphere equator along the axis $N_{MG}$ (see Fig. 1).

The brightly expressed northern-southern asymmetry of redshifts of Seyfert galaxies in Fig. 10b has a dependence shown by open circles in Fig. 11b. The closed circles represent similar points of Fig. 8, with the normalised values $<z>$ (as in Fig. 10a). In northern hemisphere of the Metagalaxy both dependencies within the limits of mistakes are in agreement with themselves. However, in a southern hemisphere a stronger latitudinal dependence $<z>$ of Seyfert galaxies in comparison with the quasars is seen.

## 4. Discussion of the results

### *4.1. On the Metagalaxy*

We will now try to obtain deeper insight into the results obtained above. For further analysis we will distinguish the concept of the Universe and the concept of the Metagalaxy. We will assume that the Metagalaxy is smaller than the Universe and is only one of its objects (in all probability, it is the largest object presently accessible to observation). As any other object, it then has a finite dimension and a finite mass. Within a different approach, the Universe and the Metagalaxy are the same, in which case there is nothing else in the world. However, it is well known that matter is structured from the micro- to the macrocosms. From general considerations it is clear that very grave causes are required for the structural staircase of matter to be completed in the region of either small or large masses—for example, the selfclosure of space around an enormously large mass. But if there is one closed world, we can also imagine yet another one, identical to it, since it is unlikely that nature produces only one specimen of anything. In view of this, we will as-

417sume that there are no reasons to restrict our world, so that the Metagalaxy is only part of the large Universe, which is infinite, in all probability.

The idea of the Metagalaxy itself is not new and is sometimes developed by researchers (see, for example, [8,14–20]). In [16,17], it was proposed to consider the Universe as that which consists of a number of "miniuniverses"; possibly, one of such miniuniverses" is precisely our Metagalaxy. In any case, this idea makes it possible, in my opinion, to develop a schematic pattern that enables one to interpret, in one way or another, the results obtained above.

Let's accept sphere with the radius of $R_{MG} \approx 10^{26}$ m $\approx 4000$ Mpc for the outer boundary of the Metagalaxy, whose mass $M_{MG}$ is approximately

$$M_{MG} \approx (4\pi/3)\,(R_{MG})^3 \rho \approx 4\times 10^{52} \quad [\text{kg}]\,, \qquad (11)$$

where $\rho \approx 10^{-26}$ kg/m$^3$ is the average matter density in the Metagalaxy [1]. At the present time, there are indications (see, for example, [8,18–22]) of some formidable structures in space. For example, the measurements of the peculiar velocities of 400 elliptic galaxies in [21] exhibit the presence of an "enormous motion" of matter. All clusters and superclusters occurring in the vicinity of our Galaxy move toward the point whose equatorial coordinates are $\alpha = 208.1°$ and $\delta = -55.5°$—this is the so-called Grand Unification point. A simulation of this gravitating mass yields an estimate of $M_A \sim 5\times 10^{16}\,M_C \approx 10^{47}$ kg, where $M_C$ is the mass of the Sun. The radius $R_A$ of the sphere of this attractor can be estimated by using the relation

$$R_A = R_{MG}\,(M_A/M_{MG})^{1/3} \approx 56 \text{ Mpc}. \qquad (12)$$

One can see that this radius is quite commensurate with the shift of the observation point with respect to the presumed center of symmetry of the volume distribution of "average quasars" [see Eq. (6)]. The attractor itself is situated approximately in the same direction (the supergalactic coordinates of its center are $l_{SG} \approx 269°$ и $b_{SG} \approx -5°$) as the Metagalaxy center. Thus, we see that the motion of matter now becomes more comprehensible − it is a motion in the field of metagalactic gravitational forces.

Further, we consider some interesting facts from astronomy that, in my opinion, favour the existence of a finite Metagalaxy. It is well known that Seyfert galaxies are, as a rule, spiral galaxies, but that they are characterised by an enhanced growth of luminosity toward the center. With respect to ordinary spiral galaxies, their number is about 1% [1]. It is astounding that the overwhelming majority of Seyfert galaxies are oriented flatwise with respect to a terrestrial observer [1]. This fact has so far remained unexplained. However, the answer is clear within the hypothesis of a Metagalaxy: the axes of rotation of Seyfert galaxies are aligned with the force lines of the gravitational field directed to the Metagalaxy center. In this case, periodic oscillations of the energy of gravitational coupling between the rotating parts of Seyfert galaxies and the gravitational field of the Metagalaxy are minimal. Moreover, the fact that the axes of rotation of Seyfert galaxies are directed toward us is yet another piece of evidence that our Galaxy is not far off the Metagalaxy center.

### *4.2. Central Asymmetry of the Abell Clusters of Galaxies*

The distribution of galaxies in space is non-uniform. Their significant part is grouped in clusters containing tens, hundred and even thousand (rich clusters) galaxies. These objects are much closer to us, than Seyfert galaxies and quasars. Probably they form the central region of the Metagalaxy. If the received above results are really caused by the certain shift of our site relative to the centre of the Metagalaxy, this shift should be also noticeable against a background of the clusters of galaxies.

Let's consider a spatial structure of rich clusters of galaxies from the Abell catalogue [23]. It is one of the widely known catalogues, which contains the richest clusters. Let's take not all objects, but only those which are nearest us, with the redshifts $0.02 \leq z \leq 0.1$. Their distribution in celestial sphere (535 objects) is represented in Fig. 12a in galactic coordinates. Some clusters form close groups caused by a large-scale cellular structure of the Metagalaxy [1]. In Fig. 12b the map of sizes (7) in units of $\Delta z = 0.005$ for the scanning circle with radius 45° is shown. The picture in many respects is similar to Fig. 7a and Fig. 10a. The "black hole" stands out here at $l_G \approx 300°$, which surpasses similar "holes" in sizes in the Figures mentioned above.

Fig. 13 shows the changes in number of these clusters (a) and their average redshifts (b) depending on the galactic longitude (with a step of $\Delta l_G = 10°$). It is seen, that the horizontal lines – average sizes – do not correspond to a real distribution of points, which correlate much better with dashed curves – approximations of the data by the function (3). In Fig. 13a, in spite of separate narrow peaks, the gradual change of average number of clusters from minimum (at $l_G \approx 135°$) up to maximum (at $l_G \approx 315°$) is observed. It is accompanied in Fig. 13b by an opposite change of average redshifts relative to $<z>_A = 0.056 \pm 0.001$, with $<z>_{max} \approx 0.065$ and $<z>_{min} \approx 0.043$ near phases of extremes mentioned above.

The results in Fig. 12b and Fig. 13 obviously contradict to the hypothesis on isotropic distribution of galaxy clusters of in environmental space (with $z \leq 0.1$). To understand the physical sense of these results, we will address to Fig. 14. Of course, this is a very rough scheme, where many proportions have been violated because of the incommensurability of scales of the Galaxy and the Metagalaxy. Nevertheless, it allows to draw the certain picture within the framework of those representations spoken above. The circle (with radius $R_{Ab} \approx 380$ Mpc) designates the external border of space volume, considered by us. Its centre ($O$) corresponds to the centre of the Metagalaxy. The location of our solar system in the Metagalaxy is specified by the letter $S$. As has been shown above (Fig. 7–11), it was displaced from the centre $O$ in the direction of the axis $N_{MG}$ − the North Pole of the Metagalaxy (see Fig. 1) − to the closest region of the clusters of galaxies. Figures designate the galactic and the supergalactic longitude. The circle with radius $<R>_{Ab} \approx$ 210 Mpc represents a sphere of an "average cluster", in Fig. 13b the horizontal line with $<z>_{Ab} = 0.056$ corresponds to it. The dashed circle represents approximation given in Fig. 13b recalculated into distances using the formulae (1) and (2). Figure on the whole reflects a relative arrangement of objects interesting for us. This sight is "open" at inter section of the Metagalaxy through its centre by a plane parallel the plane of the Galaxy.

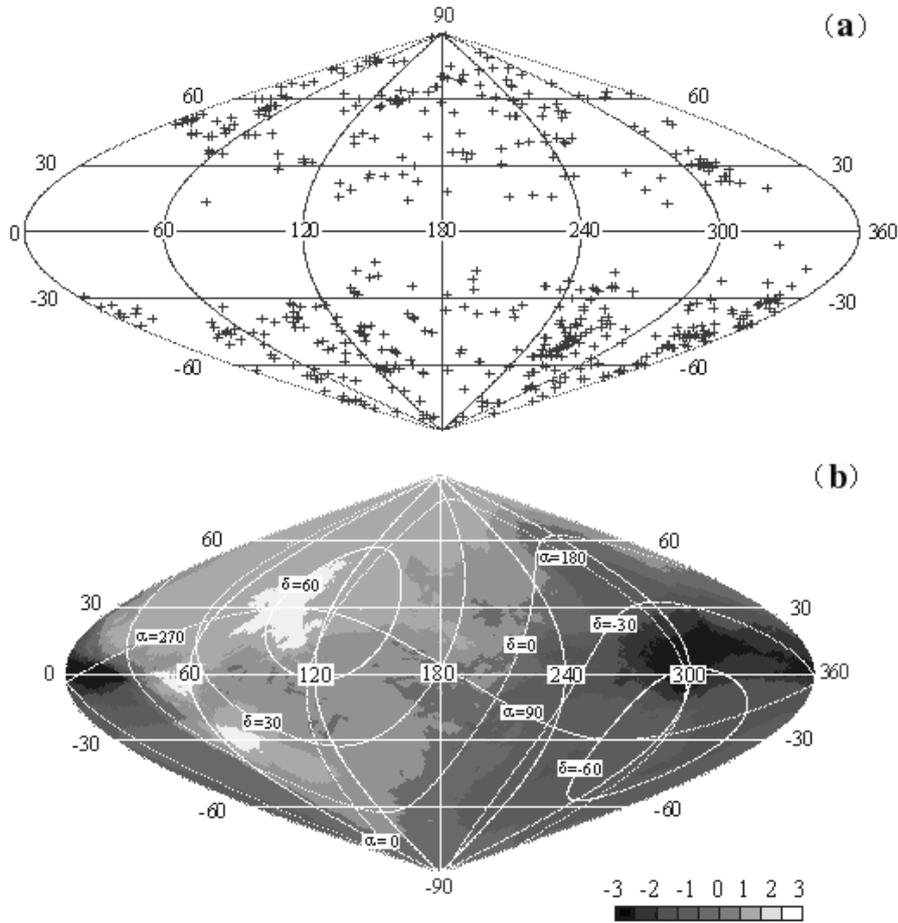

**Fig. 12.** Distributions on the developed heavenly sphere in galactic coordinates of 535 rich clusters of galaxies from the catalogue [23] from $0.02 \leq z \leq 0.1$ (a) and deviations of their average redshifts from $<z>_A = 0.056$ of the whole sample (b). The shaded scale shows limits of changes with a step $\Delta z = 0.005$

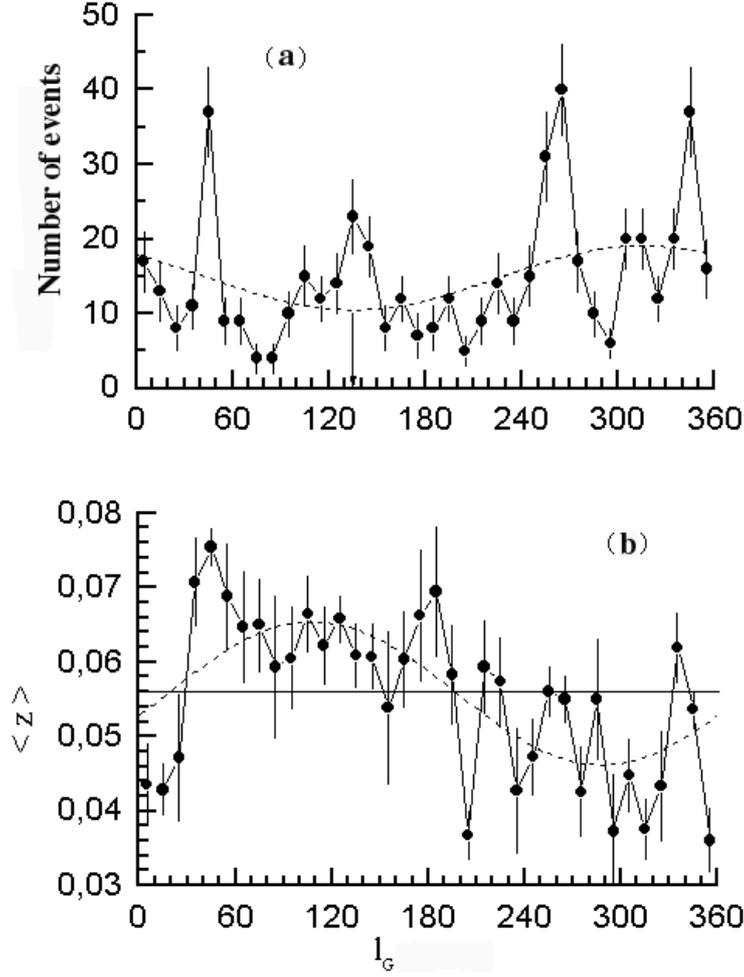

**Fig. 13.** Distributions in numbers of rich clusters of galaxies (a), submitted in Fig. 12a, and their average redshifts (b) in dependence of the galactic longitude (with a step of $\Delta l_G = 10°$). The horizontal line corresponds to average values; the dashed curves represent the best approximation of the data by the function (3).

In Fig. 14 it is seen, that the minimum number of clusters should be in the cone *1*, directed along the axis $N_{MG}$, and the maximum one is in the opposite cone *2*. At the consecutive review of the sky in galactic spherical sectors (from $l_G = 0°$ in the course of the arrow) the number of events in these sectors will vary according to the dashed curve in Fig. 13a. The piece $OS \approx \Delta R$ characterizes a shift of the point of the observer concerning the centre of the Metagalaxy. Let us now estimate it numerically in the direction of vector with equatorial coordinates $\alpha_N \approx 13°$ and $\delta_N \approx 70°$, which we shall take as a direction of the Metagalaxy North Pole. In this case in northern and southern hemispheres of the Metagalaxy there are accordingly congestion's of galaxies $N_1 = 213$ and $N_2 = 322$ respectively. Their portion $p = N_1/N_2 \approx 0.66$ is connected with a relative shift $d = \Delta R/R_{Ab}$ by the ratio

$$d(3 - d^2) = 2(1 - p)/(1 + p) , \qquad (13)$$

from which we have $d \approx 0.14$ and $\Delta R \approx 53$ Mpc – in complete conformity with the estimation (6).

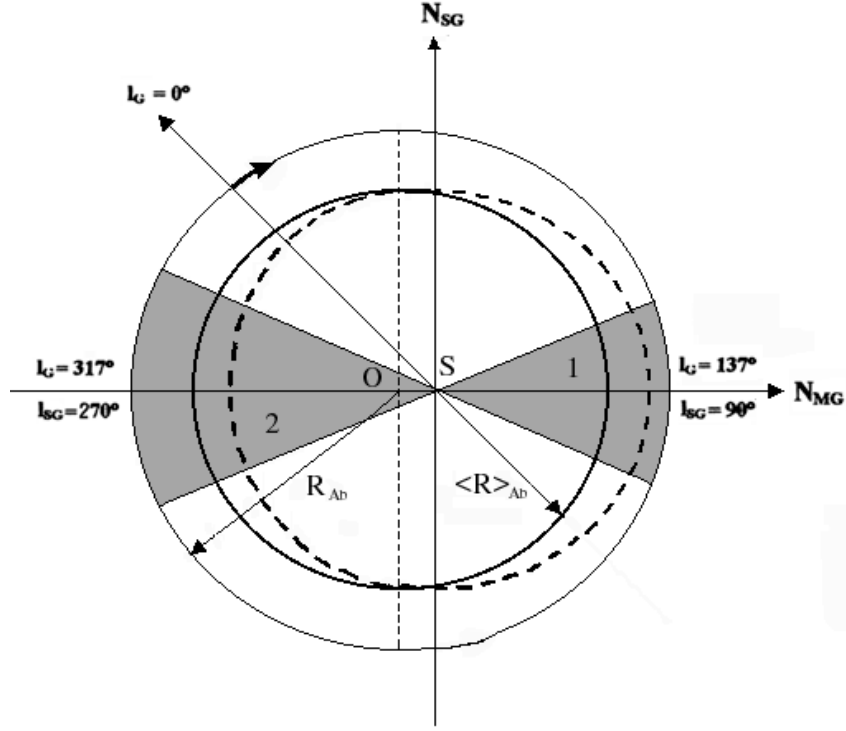

**Fig. 14.** The scheme of mutual arrangement of clusters of galaxies (the circle with radius $R_{Ab}$) and terrestrial observer ($S$) in the centre of the Metagalaxy: a circle with $<R>_{Ab} \approx 210$ Mpc corresponds to the sphere of an "average clusters"; a dashed circle corresponds to the sphere of an "average clusters", shifted along an axis $N_{MG}$ by $\approx 50$ Mpc; figures represent the galactic ($l_G$) and the supergalactic ($l_{SG}$) longitude; $N_{SG}$ shows the North Pole of the Supergalaxy.

Figure 14 also allows to understand the mechanism of formation of average redshifts of clusters of galaxies. It is seen, that the maximum of distribution (dashed curve) in Fig. 13b at $l_G \approx 130°$ is caused by the sample of events in the cone *1* lying further from the centre in comparison with the cone *2*, where at the expense of events near to the centre of the Metagalaxy with $z \approx 0$ the average redshift is of minimal size. If we recalculate the values of approximation curve in Fig. 13b by the formulae (1) and (2) into the corresponding distances and postpone them from the centre of the Metagalaxy, then the dashed circle in Fig. 14 (with the centre $S$) will be received. The radius of this circle almost coincides with $<R>_{Ab}$, but its centre is offset along the axis $N_{MG}$ by $\Delta R_{MG} \approx 42$ Mpc. This shift wonderfully coincides in size with the estimation $\Delta R_{MG} \approx 44$ Mpc received above (see chapter 2.2).

### *4.3. Gravitational Redshift*

It is accepted to consider, that the redshifts of the spectra of radiation of far galaxies are caused exclusively by Doppler effect of the extending Metagalaxy after the onset of the Big Bang. However, as is known, many characteristics of radiation also depend on the gravitational potential of environmental space. In this connection, the magnitude and the spatial distribution of the gravitational potential of the Metagalaxy are of considerable interest. In regions beyond the Metagalaxy (for $R \geq R_{MG}$), it is given by the wellknown Newton formula

$$\varphi(R) = -GM_{MG}/R, \qquad (14)$$

where $G = 6.67\times10^{-11}$ kg$^{-1}$m$^3$s$^{-2}$ is the gravitational constant. The question of distribution of the gravitational potential within the Metagalaxy is more involved. Within a uniform sphere, it is given by

$$\varphi(R) = -GM_{MG}(3 - (R/R_{MG})^2)/2. \qquad (15)$$

At $R = R_{MG}$, the two formulas match each other, yielding the same result, $\varphi_{MG} = \varphi(R_{MG}) \approx 3\times10^{16}$ m$^2$/s$^2$. At the Metagalaxy center, this potential is 1.5 times greater, the gravitational field itself is vanishing there. The gravitational potential is the work that must be performed to remove a unit mass from a given gravitational field. In order to escape from the gravitational field of the Metagalaxy, a body of mass $m$ must perform the work $A = -m\varphi_{MG}$. For this, it must have the kinetic energy $K = mv^2/2 = -A$ and, hence, the speed $v = \sqrt{-2\varphi_{MG}} \approx 2.5\times10^8$ m/s, which is close to the speed of light.

The following intriguing fact is noteworthy. The square of the speed of light, $c^2 = 9\times10^{16}$ m$^2$/s$^2$, is nearly equal to the Metagalaxy potential within the errors in the estimation of $\varphi_{MG}$. The Einstein mass–energy relation states that the total energy of a body is $E = mc^2$. It appears that this energy is approximately equal to the gravitational potential created by the entire mass of the Metagalaxy. Yanchilin [24] considered some surprising consequences from this circumstance, but we will not dwell on them here.

Already the first researchers of quasars (see, for example, [25]) have paid attention to the possible contribution into the value of redshifts of these objects of gravitational potential of the Metagalaxy. Let us now once again return to this problem. The redshift is defined as the value

$$z = (\lambda_2 - \lambda_1)/\lambda_1 \tag{16}$$

of shift of the length of wave of electromagnetic radiation $\lambda_1$ of the far object measured at the moment of radiation by the observer located there in the point *1*, relative to the length of a wave $\lambda_2$, come to the point *2*, where we are with the receiver of radiation. In this case it is supposed that our standard with the length of wave $\lambda_1$ is the same, as well as the point *1*.

Let's rewrite (16) in another kind:

$$1 + z = \lambda_2/\lambda_1 = (\nu_1/\nu_2)(c_1/c_2), \tag{17}$$

where $\nu_1 = c_1/\lambda_1$ and $\nu_2 = c_2/\lambda_2$ are frequencies of radiation; $c_1$ and $c_2$ are local speeds of light in points *1* and *2* of the Metagalaxy. Here during further discussions we shall take into account, that the local speed of light at a distance $R$ from the centre of the Metagalaxy is determined by the Einstein formula [1,25]:

$$c(R) = c_0(1 - |\Phi(R)|), \tag{18}$$

where $\Phi(R) = \varphi(R)/(c_0)^2$ is the dimensionless gravitational potential of the Metagalaxy; $c_0$ are the speeds of light far from the Metagalaxy, where its gravitational potential is equal to zero. Let's transform (17) with the account of (18) to another form which is more convenient for further analysis:

$$1 + z = (\nu_1/\nu_2)((1 - |\Phi_1|)/(1 - |\Phi_2|)). \tag{19}$$

In a general view the change of frequency $\nu_1$ of a moving source into to the observable frequency $\nu_2$ at the expense of Doppler effect occurs under the law [1]

$$\nu_1/\nu_2 = (1 - \beta_1\cos\theta)/\sqrt{1 - \beta_1^2}, \tag{20}$$

where $\theta$ is a corner between the vector of speed of the source $v_1$ ($\beta_1 = v_1/c_1$) and a direction of distribution of the wave. At $\theta > \pi/2$ the source is moved of the observer (redshift), and at $\theta < \pi/2$ is come close (violetsift). The denominator in the formula (20) characterizes a relativistic deceleration of a course of time in a moving source.

Let's rewrite (16), with the account of (19) and (20), once again as:

$$1 + z = \lambda_2/\lambda_1 = (\lambda_2/\lambda)/(\lambda/\lambda_1) = (1 + z_v)(1 + z_{gr}), \tag{21}$$

where the relation

$$(1 + z_v) = \nu_1/\nu_2 = (1 + \beta_v)/\sqrt{1 - \beta_v^2} \tag{22}$$

determines the redshift at the expense of cosmological expansion of the Metagalaxy with the relative radial speed $\beta_v$. The coefficient

$$1 + z_{gr} = (1 - |\Phi_1|)/(1 - |\Phi_2|) \tag{23}$$

reflects the additional change of the observable value (16) at the expense of gravitational potential of the Metagalaxy. From (17) and (21) we have:

$$z_v = (1 + z)/(1 + z_{gr}) - 1. \tag{24}$$

Let's estimate the value $z_v$ in the case of gravitational potential (15). In the point of observation *2* near the centre of the Metagalaxy we have $c_2^2 = 9 \times 10^{16}$ m²/s² and $\Phi_2 \approx 1.5\varphi_{MG}/c_2^2 \approx 0.5$. For the quasars located near its external boundary we have $c_1 \approx 1,5c_2$, $\varphi_{MG} \approx 3 \times 10^{16}$ m²/s² and $\Phi_1 \approx \varphi_{MG}/c_1^2 \approx 0.148$. From (23) we find the correction $1 + z_{gr} \approx 1.7$. It puts the Doppler redshift $z_v \approx 0.52$ in correspondence with the value obtained above $<z>_Q = 1.586$. It decreases an estimation of radius of sphere of "average quasars" up to $<R> \approx 1584$ Mpc, that is by $\approx 1.9$ times.

The potential (15) hardly reflects a real picture in the Metagalaxy. It is, probably more suitable, for a gravitational field inside the Sun and other objects of the same type. And the matter in the Metagalaxy, similarly to spherical congestion's of stars, galaxies and superclusters of galaxies, has, most likely, a growing concentration from the periphery to the centre. With such a distribution of matter the correction (23) can reach values 2–3 and more, and it will lead to the decrease of observable redshift of quasars by $\approx$ (6–10) times.

Such a change of the redshift in many respects explains the extraordinary powerful radiation of the quasars in the wide range of frequencies (from $10^{12}$ up to $10^{22}$ Hz), which does not fit the general law of "distance – luminosity" for other objects of the Metagalaxy. It is seen in Fig. 15 taken by us from [1], where the quasars are represented in the right top corner by crosses. The di-

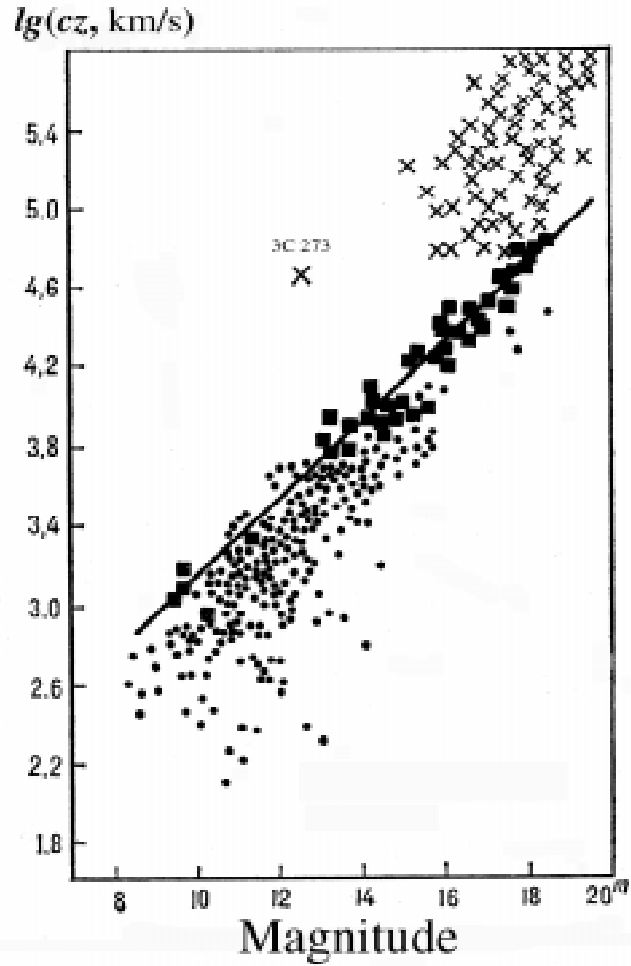

**Fig. 15.** The Hubble diagram from [1], where the redshifts of quasars (×), the brightest galaxies in clusters (■) and galaxies of a field (●) are compared with their visible star magnitude. The direct line corresponds to the dependence (24).

$$m = 5\log(z) + \text{const}, \tag{25}$$

rect line corresponds to the dependence found by Hubble for 29 galaxies located at distances from us ≤ 2 Mpc. The brightest Seyfert galaxy in the sky 3C 273 has $z = 0.158$. The account of the contribution of gravitational potential of the Metagalaxy, about which was spoken above, displaces the redshift of quasars by the scale $\log(cz)$ downwards almost by a factor of 10 and eliminates the available contradiction.

Apparently by the local change of gravitational potential one can also explain the occurrence of "black holes" in Figs. 7, 9 and 12 near the poles of the Metagalaxy, which are characterized by a fast decrease of value $<z>$ in Fig. 8 and Fig. 10b as the size of rings around the poles decreases. Actually these "holes" lay on the line of visible intersection of planes of the Galaxy and Superga-laxy (see Fig. 1), along which the equatorial regions form the maximum total dipole gravitational field. It, probably, increases the energy of radiation, coming from outside of these structures, and reduces its primary redshift, and at the most – in the directions of "black holes".

It is possible, that the weight of the attractor located in the centre of the Metagalaxy plays the additional role in this process. It also increases the energy coming from quasars and Seyfert galaxies of radiation and reduces its redshift at the expense of its attraction. By virtue of the mentioned asymmetry of our location relative to the centre of the Metagalaxy this effect is, probably, manifested as much as possible just along the line passing through the centre of the Metagalaxy and the observer.

### *4.4 On Possible Rotation of the Metagalaxy*

The idea about that the Metagalaxy along with the expansion has a global rotation, has repeatedly attracted attention of the researchers to itself [18,27–33]. From the ratio of (20) it is seen, that even in the case, when a distance between a source and observer does not vary ($\theta = \pi/2$), the occurrence of the redshift at the expense of, so-called, transverse Doppler effect is possible:

$$(1 + z_\varpi) = 1/\sqrt{1 - \beta_\varpi^2}, \tag{26}$$

connected with the rotation of source, where $\beta_\varpi = v_\varpi/c$ and

$$v_\varpi = \omega R_\varpi, \tag{27}$$

$v_\varpi$ – the rate of motion of the source around a circle with radius $R_\omega$. The condition follows from the equality of relationship (22) and (26)

$$\beta_\varpi = 1.41/\sqrt{1 + 1/\beta_v}, \tag{28}$$

at which the relative rate of rotation $\beta_\varpi$ and radial movement $\beta_v$ result in the identical redshifts. The meanings of these values $\beta_\varpi$ and $\beta_v$ depending on $z$ are shown in Fig. 16a by dashed and solid curves respectively. It is seen that at small $z$ the source of radiation should have a higher rate of motion around a circle, than a radial one, in order the observable signals would have the identical redshift in both cases. However, in the limiting relativistic case this distinction completely disappears.

If we rewrite the Hubble law (1) in another form:

$$v = HR_v, \tag{29}$$

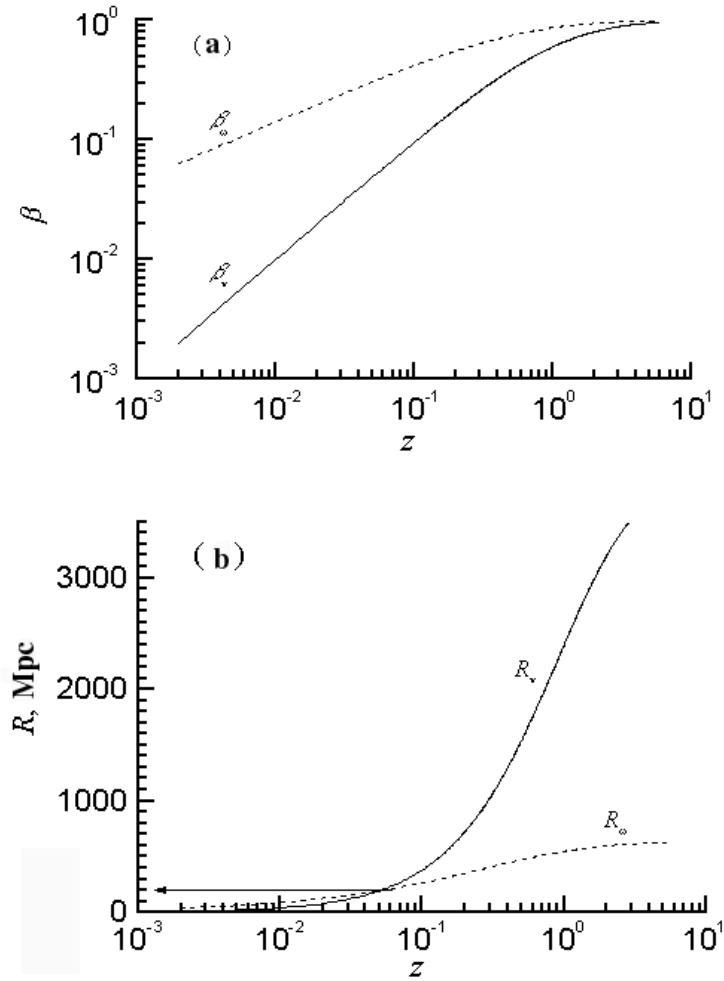

**Fig. 16.** Relative speeds of radial movement $\beta_v$ and rotation $\beta_\varpi$ (a), and distances, appropriate to them, up to the excluded ($R_v$) and rotating ($R_\varpi$) of sources of radiation (b), leading to identical observable redshifts.

where the distance $R_v$ is measured in units of 1 Mpc, and compare it with the equation (27) for the tangential rate of rotation of a body, then it is easy to notice, that the Hubbles law (29) can be interpreted not only as the expansion of the Metagalaxy with the increased radial velocity (at each length 1 Mpc ($\approx 3\times10^{19}$ km) it increases by 75 km/s), but also as its rotation with the angular velocity $\omega = H = 1/T$, equal to one revolution for $T \approx 13$ billion years.

Let's consider a possible rotation of the Metagalaxy a bit of more widely. In Fig. 16b the dashed curve shows the distances $R_\varpi$ from an axis up to points of the circle, which moves with a velocity

$$v_\varpi = c\beta_\varpi = c\sqrt{1 - (1+z_\omega)^{-2}} \tag{30}$$

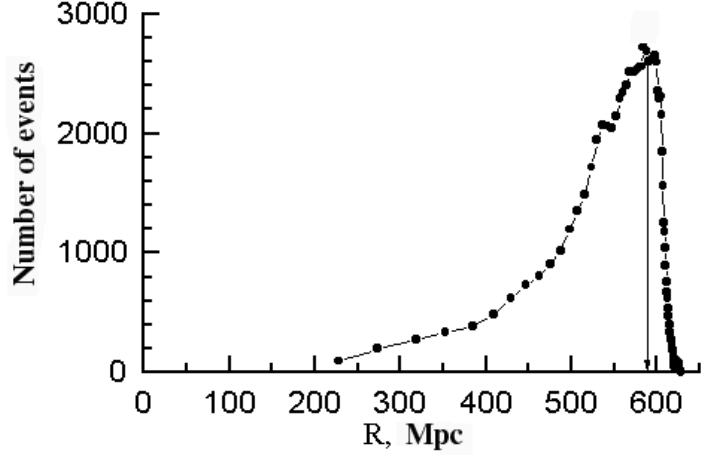

**Fig. 17.** Distribution in numbers of quasars from the catalogue [12] depending on their distance up to the centre of the Metagalaxy, rotating with angular speed $\omega = H \approx 75$ (km/s)/Mpc; arrow indicate the radius of spheres of "average quasars".

at the expense of the angular rate of rotation $\omega$. The solid curve in Fig. 16b shows distances (1), corresponding to the equation (22). It is seen, that in the first case, when the Metagalaxy rotates as a quasifirm body, the distances between its objects with redshifts $z$ at $R_\omega < 200$ Mpc should be more, than it follows from (1) and (2), and at $R_\omega > 200$ Mpc, on the contrary, they should be less. Thus, the maximum size of the rotating Metagalaxy

$$R_{MG} = c/(2\pi H) \approx 637 \text{ Mpc} \tag{31}$$

becomes much less, than extending ($\approx 4000$ Mpc) one. In such the Metagalaxy the quasars are located more densely in a relative narrow spherical layer. It is seen in Fig. 17, where in the external layer at $R \geq 500$ Mpc there are $\approx 94\%$ of all quasars [12], and the arrow specifies the radius of sphere of the "average quasars" with $<z>_Q = 1.586$.

Let's note one curious opportunity connected with the rotation of the Metagalaxy. In Fig. 7b the light and dark triangles show directions of the maximum deviation of temperature of cosmic microwave background (CMB) radiation from an average level all over the whole sky ($\approx 2.7$ K) by $+3.5$ and $-3.5$ mK, respectively [1]. This anisotropy has a dipole character and is caused by a movement of the Sun relative to CMB toward the Lion constellation with the equatorial coordinates $\alpha \approx 165°$ and $\delta \approx 8°$ (supergalactic coordinates $l_{SG} \approx 190°$ and $b_{SG} \approx -28°$) with the velocity $\approx 600$ km/s [34]. At the expense of rotation of the Metagalaxy such a speed will be observed at a distance from an axis $R_\varpi \approx 600/75 \approx 8$ Mpc, which roughly does not contradict to the estimation

$$R = \sqrt{(\Delta R_G)^2 + (\Delta R_{SG})^2} \approx 22 \text{ Mpc} \tag{32}$$

of this distance from the relationship (6). If we assume, that the rotation of the Metagalaxy occurs counter-clockwise relative to its North Pole (see Fig. 1), then in this case the velocity vector (29) will be directed to the point with coordinates $l_{SG} \approx 180°$ and $b_{SG} \approx -65°$, which roughly indicates to the side of the maximum CMB temperature. It speaks that the anisotropy of CMB can be caused by the rotation of the Metagalaxy with the angular velocity mentioned above.

Let's notice, that the different values of $R_\varpi$ and $R$, and also appreciably differing directions of the velocity vector (29) and the pole of heat of CMB in many respects are caused by the inaccuracy of $\Delta R_G$ and $\Delta R_{SG}$ values found above, which we give here only for demonstration of one more opportunity for the decision of the anisotropy problem of CMB.

The rotation of the Metagalaxy allows to give one more interpretation of the phenomenon of "black holes" in Fig. 7 and Fig. 10. Their occurrence near poles of the Metagalaxy can be connected with the fact that objects located near the axis of rotation of the Metagalaxy have no a noticeable redshift. It is seen in Fig. 18, where the curve *2* shows values

$$z_\varpi(b_{MG}) = 1/\sqrt{1 - \beta_\varpi^2} - 1 \tag{33}$$

depending on the metagalactic latitude $b_{MG}$ of the rotating sphere with radius (31) and

$$\beta_\varpi = 0.92\cos(b_{MG}). \tag{34}$$

Together with the isotropic gravitational redshift (the horizontal line *1*) it results in the redshift shown by the curve *3*, according to (21). It is in agreement with the data of observations at $|b_{MG}| \geq 45°$, but in the equatorial region of rotating Metagalaxy much greater values of *z* are expected in this case. The disagreement appreciably decreases if we enter into account not a local light velocity $c_2 = 300000$ km/s in the observation point *2* near the Metagalaxy centre, but $c_1 \approx 1.5c_2$ for the quasars located near its external border. In this case the angular rate of rotation of quasars in (33) becomes less by a factor of $\approx 1.5$ and the curve *3* in Fig. 18 turns into the dashed curve *4*, which agrees much better with the results of observations.

It should be mentioned, that we did not pose a task to achieve a good agreement of calculations in Fig. 18 with the observed distribution. At the present stage it is difficult to achieve it because much knowledge about the Metagalaxy is not yet known. The main sense of the curve *4* is that on this way there can be answers to the phenomenon of "black holes" in Fig. 7 and Fig.10 in principle.

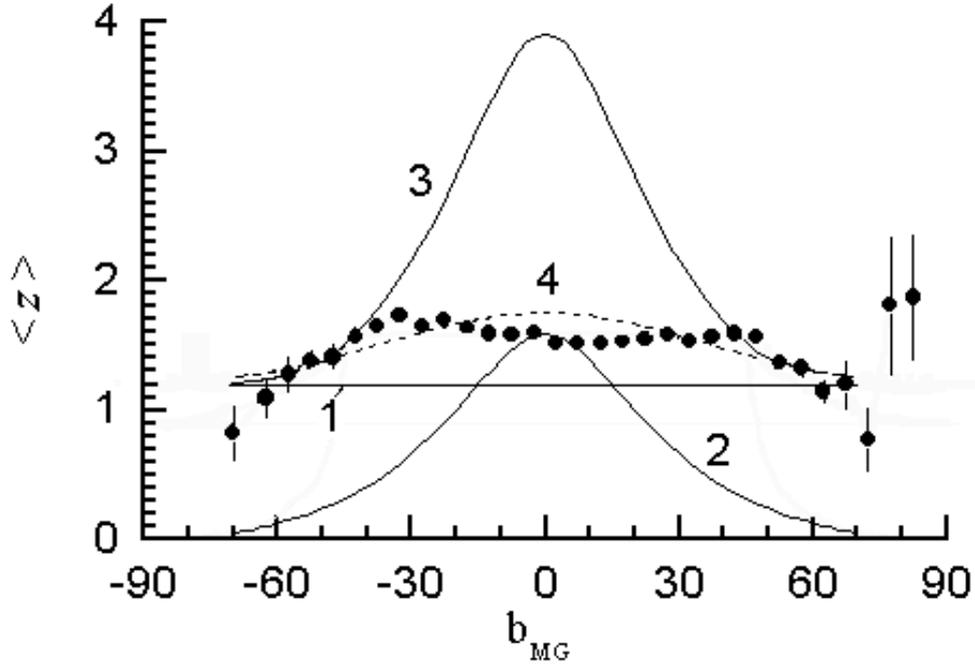

**Fig. 18.** The redshifts of quasars in different models of the Metagalaxy depending on its latitude: ● shows the average values submitted in Fig. 8; *1* represents one of the possible variants of gravitational redshifts; *2* represents cross Doppler effect (33) for the rotating sphere with radius $R_{MG} \approx 637$ Mpc and $\beta_\varpi = 0.92\cos(b_{MG})$; *3* represents the joint action of factors *1* and *2*; *4* represents the variant *3* for speed of light $c_1 \approx 1.5c \approx 450000$ km/s from near the border of the Metagalaxy.

Let's additionally cite some experimental facts testifying on a possible rotation of the Metagalaxy. In [18] the anisotropy of radioradiation of 132 objects practically covering all the celestial sphere was analyzed, and the general direction of their rotation in galactic coordinates was found

$$b_G = 24° \pm 20°, \quad l_G = 295° \pm 25°, \tag{35}$$

for the value

$$\omega/H = 1.8 \pm 0.8, \tag{36}$$

where $H \approx 50$ (km/s)/Mpc. It is wonderful, what even the rough estimation (35) correlates well with a direction of large-scale anisotropy of the Metagalaxy observable in a form of a "black hole" of the minimum redshifts in Fig. 12b.

The latest data [35,36] on ray velocities and distances of 180 galaxies of the Local volume indicate to the significant deviation from isotropic expansion of the Metagalaxy. The observable distribution of the local size of Hubble parameter can be submitted by the three-axial ellipsoid, having the relationship of axes $H_a:H_b:H_c = 81:62:48$ (km/s)/Mpc with an error ~ 8%. The minimum value ($H_c$) is observed along the polar axis of the Supergalaxy (axis $N_{SG}$ in Fig. 1), and the region of the greatest value is in the plane of equator of the Supergalaxy forming the corner 29° ± 5° relative to the centre of supercluster of galaxies Virgo (axis $N_G$ in Fig. 1). In the whole the local picture of non-hubble movements of galaxies poorly corresponds to the known model [37] which supposes that galaxies symmetrically moves to the centre of the Virgo.

We believe, that it, most likely, speaks about the rotation of galaxies around an axis focused approximately along an axis $N_{MG}$ (see Fig. 1). In this case components $H_a$ and $H_c$ form a vector located approximately in the plane of the Metagalaxy, and directed under a corner ≈ 30° to the plane of the Supergalaxy. If the rotation occurs counter-clockwise, this vector indicates to the side of maximum temperature of the CMB.

The Local group of galaxies has presented one more surprise, which consists in the following. On the one hand, velocities of 20 galaxies-dwarfs located from us not further than 3 Mpc, have a surprisingly small spread around the linear law (1) (only 25 km/s [35,36]) in the system of the centre of mass of the Local group. On the other hand, as was told above, the centre of masses of this group moves relative to CMB with the velocity ≈ 600 km/s. In dynamics controlled by the gravitation of galaxies themselves, such a picture is absolutely not possible. The computer modeling [38] has shown, that the dispersion of velocities should be greater by a factor of 3–5. For an explanation of this phenomenon in [39] it is offered to consider the pressure of space vacuum in addition.

However, the decision of the specified problem is possible also within the framework of the scenario of the rotating Metagalaxy. In this case the Local volume will move in a circle with the radius $R_\varpi \approx 8$ Mpc as a single whole with the velocity ≈ 600 km/s, and the dispersion of velocities of separate galaxies

$$\Delta v_\varpi = \omega \Delta R_\varpi = H \Delta R_\varpi \approx 75 \cdot 0.335 \approx 25 \text{ km/s} \qquad (37)$$

will be determined by the average remoteness of the Galaxy and the Andromeda (distance between them is ≈ 0.67 Mpc) from the common centre of mass for the whole group. Both these galaxies have a total mass $M_{LG} \approx 1.5 \times 10^{12}$ $M_C \approx 3 \times 10^{43}$ kg being the most important members of the Local group, around of which the majority of other members is concentrated [1,35,36].

## 5. Conclusion

The analysis of three-dimensional spatial distribution of quasars with redshifts $z \leq 6$ has allowed to establish the following. Although the catalogue [12], containing 48921 quasars, is for now extremely non-uniform in density of the distribution of objects on the celestial sphere (see Fig. 2), but nevertheless allows to receive a sufficiently correct topological picture of their average redshifts (see Fig. 3−5,7,8) reflecting a distribution on the remoteness from the observer (see Fig. 6) of the farthest and mysterious objects of the Universe.

In the arrangement of quasars there is a global anisotropy, caused by the shift a place of the observer from the centre of sphere with the radius $\approx 2960$ Mpc towards of the vector with equatorial coordinates $\alpha \approx 13°$ and $\delta \approx 70°$ on $\approx 50$ Mpc (see chapter 2). In the direction of this vector and opposite to it there are extensive regions (in the volume of corporal corners $\Omega \approx 2\pi(1-\cos 40°)$), where the gradual decrease of average redshifts around specified vectors up to the minimum value (see Fig. 7 and Fig. 8) is observed. The received results are not connected with any methodical features of the catalogue [12], and reflect the certain structure of distribution of substance in volume accessible maximum ally for observation of environmental space.

Similar global anisotropy is observed in spatial distribution of Seyfert galaxies (see Fig. 10 and Fig. 11) and rich clusters of galaxies (see Fig. 12 and Fig.13). This anisotropy is indicative of the existence of the Metagalaxy, a gravitationally bound set of matter in the Universe within which we live and beyond which our sight can not penetrate in principle. In this case, the Galaxy is, probably, near the centre of the Metagalaxy (with coordinates of its centre $\alpha \approx 193°$ and $\delta \approx -70°$). The mathematical modeling (see Fig. 9 and Fig. 14) has shown, that the assumption of such structure of the Metagalaxy can quite explain the whole set of the results received above.

The limited sizes and mass of the Metagalaxy require to take into account the essential contribution of its gravitational potential into observable redshifts of quasars. It can explain in many respects the extraordinary powerful radiation of quasars, which is not kept within the common Hubble law (see Fig. 15) for other objects. In this case the size of the Metagalaxy becomes appreciably less, than it follows from the law (1) in the scenario of only cosmological expansion of the Metagalaxy after the onset of the Big Bang.

The received results suppose an opportunity of rotation of the Metagalaxy (see Fig. 16 and Fig. 18) around of an axis with the coordinates $\alpha \approx 13°$ and $\delta \approx 70°$ mentioned above. In this case its size can become even less (see Fig. 17). Kinematics of movement of the Local group of galaxies does not contradict the rotation with angular velocity $\omega = H \approx 75$ (km/s)/Mpc $\approx 4.8 \times 10^{-10}$ glad/year.

It is possible that the global anisotropy of redshifts of the objects considered by us is caused in one measure or another by all three cosmological mechanisms: gravitational, rotation and expansion of the Metagalaxy. The received results, undoubtedly, require the further deep and detailed research. This work will be continued.

## Acknowledgments

This work was carried out under the financial support of the Yakutsk EAS array by the Ministry of Science of Russia (reg. no. 01-30), which was included into the "List of Unique Research and Experimental Facilities of National Significance" and was supported by the Russian Foundation for Basic Research, project no. 05-02-17857a.